\documentclass[runningheads]{comsis2}

\def\journalissue{Computer Science and Information Systems 00(0):0000--0000}
\def\paperidnum{https://doi.org/10.2298/CSIS123456789X                  }
\usepackage[pagewise]{lineno}

\usepackage{longtable}
\usepackage{framed}
\usepackage{graphicx}
\usepackage{multirow}
\usepackage{tabularx}

\title{Agile Culture Clash: Unveiling Challenges in Cultivating an Agile Mindset in Organizations

\footnote{This paper is an extended version of our conference paper titled \textit{Which Challenges Do Exist With Agile Culture in Practice?}~\cite{Kuchel.2023}. In this paper, we updated the related work and added an in-depth presentation of the sample of our study. Furthermore, we provide a detailed discussion of each key challenge including a clustering of the challenges related to the dimensions of being and doing agile. Finally, we explain in more detail our conceptual model and present key take aways for practitioners and researchers.}}

\titlerunning{Agile Culture Clash: Unveiling Challenges in Cultivating an Agile Mindset in Organizations}

\titlerunning{Unveiling Challenges in Cultivating an Agile Mindset in Organizations}

\author{Michael Neumann\inst{1} 
\and Thorben Kuchel\inst{2}
\and Philipp Diebold\inst{2,4}
\and Eva-Maria Schön\inst{3}}

\institute{University of Applied Sciences and Arts Hannover\\
  Dpt. of Business Information Systems\\
  Hannover, Germany\\
  \email{michael.neumann@hs-hannover.de}
  \and
Bagilstein GmbH\\
  Mainz, Germany\\
  \email{ \{firstname.lastname\}@bagilstein.de}
  \and
  University of Applied Sciences Emden/Leer\\
  Dpt. of Economics\\
  Emden, Germany\\
  \email{eva-maria.schoen@hs-emden-leer.de}
  \and
  IU International University\\
  Frankfurt, Germany\\
  \email{phillip.diebold@iu.org}
  }

\begin{document}

\maketitle

\begin{abstract}
\textit{Context:} In agile transformations, there are many challenges such as alignment between agile practices and the organizational goals and strategies or issues with shifts in how work is organized and executed. One very important challenge but less considered and treated in research are cultural challenges associated with an agile mindset. Although research shows that cultural clashes and general organizational resistance to change are part of the most significant agile adoption barriers. 
\textit{Objective:} We identify challenges that arise from the interplay between agile culture and organizational culture. In doing so, we tackle this field and come up with important contributions for further research regarding a problem that practitioners face today.   
\textit{Method:} This is done with a mixed-method research approach. First, we gathered qualitative data among our network of agile practitioners and derived in sum 15 challenges with agile culture. Then, we conducted quantitative data by means of a questionnaire study with 92 participants.     
\textit{Results:} We identified 7 key challenges out of the 15 challenges with agile culture. These key challenges refer to the technical agility (doing agile) and the cultural agility (being agile). The results are presented in type of a conceptual model named the Agile Cultural Challenges (ACuCa).
\textit{Conclusion:} Based on our results, we started deriving future work aspects to do more detailed research on the topic of cultural challenges while transitioning or using agile methods in software development and beyond.

\vspace{6pt}\textbf{Keywords:} Agile software development, agile methods, agile culture, agile mindset, organizational culture, challenges, barriers, conceptual model
\end{abstract}

\section{Introduction}
\label{Sec1:Introduction}
Agile methods have been used in software development for more than 20 years~\cite{Abrahamsson.2002} and are widely adopted to respond to rapidly changing market conditions~\cite{Bennett.2014}. With an iterative approach and emphasis on collaboration, communication and interaction, agile methods allow for adequate responses to changing requirements and individual customer needs, enabling companies to remain competitive even in volatile markets. Nowadays, the term VUCA\footnote{The VUCA acronym was introduced by the U.S. military and describes markets that are characterized by highly dynamic conditions and thus, rapidly changing environments like customer needs and requirements.~\cite{Bennett.2014}} (volatility, uncertainty, complexity, and ambiguity) is used to characterize this context~\cite{Bennett.2014}. However, values and principles are of high importance when introducing and using agile methods in practice due to their focus on social aspects such as communication, interaction, and stakeholder integration~\cite{Chow.2008}. The Agile Manifesto~\cite{Beck.2001} provides a ground for a value-based approach to product development. Furthermore, the guidelines for specific agile methods such as Scrum~\cite{Schwaber.2020} or Kanban~\cite{Anderson2010} define further and specific values for these approaches.

Many companies face various challenges when transitioning to agile methods~\cite{Strode.2009}. A successful agile transformation depends on several factors, including the involvement of stakeholders~\cite{Hoda.2011}, correct application of agile practices across different organizational levels~\cite{Neumann.2022}, and cultural aspects of the organization~\cite{Diebold.2015}. The State of Agile Report~\cite{VersionOne.2023} identifies contradictions between organizational culture and agile values as a core challenge in agile transition. For example, actively promoting the self-organization of agile teams requires a culture that values transparency and critical thinking regarding the agile approach in use~\cite{Schoen.2015}. However, challenges arising from cultural aspects when introducing and applying agile methods in practice are diverse~\cite{Smite.2021}. Thus, contextual differences, such as regions, industries, or organizational levels, should be considered when studying agile methods~\cite{Gelmis.2022,Neumann.2023}.

One well-known distinction in applying agile is between doing and being agile~\cite{Kuepper.2017,Sidky.2007}. Doing agile (technical agility) involves the methodological application of agile methods and practices, while being agile (cultural agility) emphasizes an agile mindset based on underlying values and principles. The interplay between these two dimensions is of high relevance for a successful agile transformation and provides a starting point for an understanding of an agile culture independent from the organizational context. Thus, we define agile culture as follows:
\begin{framed}
\textbf{Definition \textit{Agile Culture~\cite{Kuchel.2023}}}: An Agile Culture reflects the behavior of people working in an organization using agile practices based on the underlying values and principles defined in the Agile Manifesto and the guidelines of agile methods.
\end{framed}

Literature presents many studies regarding the application of technical agility~\cite{Campanelli.2015,Hoda.2017}. However, agile culture has been underrepresented in the recent literature (e.g., \cite{Gelmis.2022}). To address this gap, our mixed-method study aims to identify key challenges associated with the interplay between agile culture and organizational culture. 
\newpage Hence, this paper addresses the following research questions: 
\begin{itemize}
    \item \textbf{RQ1:} \textit{What are the key challenges with agile culture?}\\
    As described above, literature show various challenges with regard to agile transformations or the use of agile methods in practice. However, we identified a lack of understanding which challenges relate to an agile culture. Thus, we want to fill this gap by answering this research question by identifying key challenges with an agile culture. 
    
    \item \textbf{RQ2:} \textit{How can we describe the relationships of the key challenges with agile culture in a systematic manner?}\\
    We know that identifying and analyzing challenges can be complex as the challenges may have interrelations with each other. Thus, this second research questions aims to provide a systematic description of the relationships between the identified challenges. 
    
\end{itemize}

The paper at hand is structured as follows: In Section~\ref{Sec2:RelWork}, we outline the related work. Section~\ref{Sec3:ResearchDesign} explains our mixed method research approach with qualitative and quantitative data. Section~\ref{Sec4:Results} presents our identified key challenges with agile culture and categorizes them to technical agility and cultural agility. Then, Section~\ref{Sec5:Model} presents our conceptual model that shows the relationships between the key challenges with agile culture. Section~\ref{Sec6:Discussion} discusses the implications on practitioners and researchers and presents the limitations of our study. Finally, the paper closes with Section~\ref{Sec7:ConclusionAndFutureWork}, a conclusion of this work and future research directions.

\section{Related Work}
\label{Sec2:RelWork}
We searched for both, primary and secondary studies in order to provide an overview of the literature related to key challenges of cultural aspects while transitioning or using agile methods or practices. The search was performed using Google Scholar. We argue the choice to use Google Scholar by considering diverse publishers (e.g. ACM, IEEE) and the high consistency of results with digital libraries like Scopus~\cite{Yasin.2020}. The literature review was designed according to the guidelines for rapid reviews by Cartaxo et al. \cite{Cartaxo2020}. We focused our literature search on agile cultural related aspects.  Table~\ref{tab1:OverviewofRelWork} gives a brief overview of the identified related work.

\begin{table*}
 \caption{Overview of the related work in accordance with~\cite{Kuchel.2023}}
  \label{tab1:OverviewofRelWork}
  \begin{tabular}{cp{3cm}p{7cm}}
\hline
Reference & Cultural level(s) & Findings \\
\hline
Neumann et al. (2023)~\cite{Neumann.2023} & National culture \& organizational culture & The authors present a causal model describing the impact of cultural aspects on agile practices. They focus on national culture using the six cultural dimensions by Hofstede and the two dimensions from the Competing Values Model. The hypothetical causal impact is described systematically per each cultural dimension on one agile practice. In total, their model presents 384 possible relationships of cultural impact on agile practices.\\
\hline
Gelmis et al. (2022)~\cite{Gelmis.2022} & National culture & Hierarchical cultural structures, which are described for Turkish culture, are challenging the use and transition of agile methods. The authors recommend transforming to a more flat organizational structure, which supports the use of agile methods and should optimize the performance of agile teams.\\
\hline
Gupta et al. (2019)~\cite{Gupta.2019} & Organizational culture & The results of the study show that a hierarchical culture hinders the use of social and technical agile practices. Development cultures support both social and technical agile practices. The authors recommend considering the underlying cultures in the organization before starting an agile transformation and describe a six-step process according to Cameron and Quinn~\cite{Cameron.2006}.\\
\hline
Iivari and Iivari (2011)~\cite{Iivari.2011} & Organizational culture & Based on an empirical study, the authors point out that a hierarchical culture is incompatible with the use of agile methods. For example, it is shown that the transition and use of agile methods in hierarchical organizational cultures leads to a steadily increasing formalization of the agile method and that it can lose key elements of agile work over time.\\
\hline
Siakas and Siakas (2007)~\cite{Siakas.2007} & National \& organizational culture & In their work, the authors describe an agile culture based on well-known success factors such as user satisfaction and stakeholder involvement. They present the connections between organizational and cultural aspects and elements of agile working such as practices and roles. The authors also point out the importance of employee motivation and dynamics and draw parallels to established frameworks such as ETHICS \cite{Mumford.2000} or TQM \cite{Deming.1986}.\\
\hline
Strode et al. (2009)~\cite{Strode.2009} & Organizational culture & Based on their empirical study, the authors present specific correlations between organizational culture characteristics and the successful application of agile methods. They point to the importance of positive evaluation of feedback and learning, a trustful social interaction, a collaborative cooperation and focusing on results.\\
\hline
Strode et al. (2022)~\cite{Strode.2022} & Organizational culture & The authors discuss the challenges of transforming into an agile organization, which involves significant changes in strategy, structure, culture, operations, and technology. These transformations led to various tensions, of which 13 are highlighted. The study demonstrates that such tensions arise not only in fully agile organizations but also during the transformation process.\\
\hline
Tolfo et al. (2011)~\cite{Tolfo.2011} & Organizational culture & The authors address the supporting and hindering influences of organizational culture on the transition of agile methods. A superficial examination of these influences can be a hindrance and lead to incorrect measures. Rather, it is important to consider the different levels of organizational culture in order to be able to develop an agile culture in the long term.\\
\hline
\end{tabular}
\end{table*}

Various authors point to the multiple facets of culture~\cite{Gupta.2019,Neumann.2023,Siakas.2007,Smite.2020}. The overview of the related work also shows that different cultural levels (country, regions, companies, departments, teams, and individuals) may have different influences on the transition and use of agile methods and practices. In addition, agile methods can be very diverse in practice due to the tailoring of the respective method~\cite{Diebold.2014,Neumann.2022}. Most of the articles we identified address the influence of organizational culture on the transition or use of agile methods and practices. Interestingly, different levels of detail are considered here. While some use organizational culture and its characteristics on an abstract level~\cite{Iivari.2011,Siakas.2007}, others try to make a specific reference to teams and individuals~\cite{Tolfo.2011}.

However, it is worth mentioning that most of the identified studies were published a decade ago. Agile has evolved since these studies were published. The diversity of agile practices and methods has grown steadily. Furthermore, the challenges in organizations have changed due to factors such as increasing digitization, volatile markets, and the Covid-19 pandemic~\cite{Neumann.2022a}. 

Although we tried to find related work dealing with key challenges of cultural aspects while using agile methods and practices, we could not identify literature dealing closely with the aim of our study. We identified some additional papers that deal with cultural facets or an agile mindset. However, we did not identify these papers as related work because they are not strongly related to our main topic, particularly the agile cultural aspect. So, to the best of our knowledge, this is the first publication that identifies key challenges with agile culture. 

\section{Research Design}
\label{Sec3:ResearchDesign}
This paper aims to identify challenges that arise from the interplay between agile culture and organizational culture. 

Based on the aforementioned aim, we formulate the following research questions: 
\begin{itemize}
    \item \textbf{RQ1:} What are the key challenges with agile culture?
    \item \textbf{RQ2:} How can we describe the relationships of the key challenges with agile culture in a systematic manner?
\end{itemize}

\begin{figure}
\caption{Overview mixed-method research approach~\cite{Kuchel.2023}}
\label{Figure1}
\includegraphics[scale=0.45]{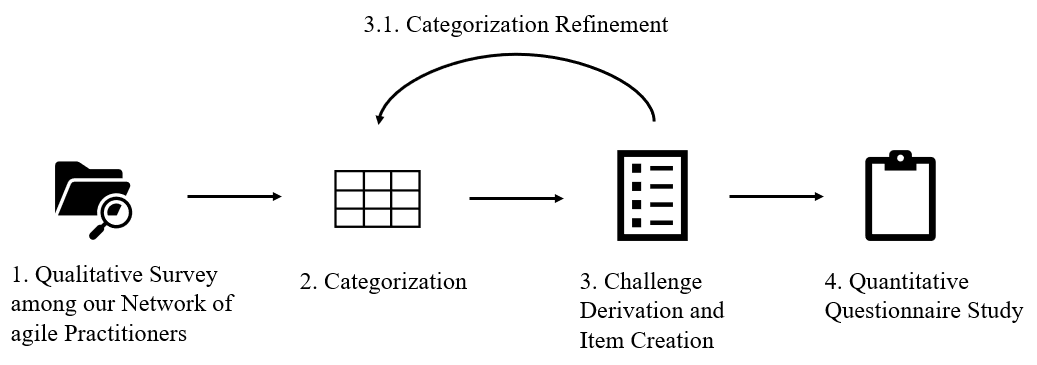}
\end{figure}

RQ1 is answered with a mixed-method approach (see Fig.~\ref{Figure1}), where we collect qualitative and quantitative data and derive the key challenges. At each step of this mixed-method research approach, we gathered feedback from practitioners and discussed it extensively among the group of authors. Based on this, we develop the ACuCa model to answer our RQ2. The ACuCa model shows how the identified key challenges are interrelated. In the subsequent discussion, we show how practitioners can apply the key challenges and the ACuCa model in practice to mitigate problems with the agile culture.

\paragraph{Step 1:} In the beginning, we started with a qualitative survey among our network of agile practitioners in order to gather input for our quantitative questionnaire. Therefore, the third author of the paper asked the question on the professional network LinkedIn: “What do you see as tangible problems with agile and culture?” We discussed how to phrase this questions in such a manner that our target group understands it and we receive reasonable data for our study. With this regard, we tried to avoid the term “agile culture” since this opens another field of discussions, that was not our focus in this research. In sum, we received 34 qualitative answers from 28 participants.   

\paragraph{Step 2:} We analyzed the data with a qualitative context analysis according to Hsieh and Shannon~\cite{Hsieh.2005}. We used a combination of directed and conventional context analysis. For the former one, we used the research literature on Agile, and the latter one complemented the analysis since research on agile culture is rather limited. The data was categorized by a coding scheme, which ended up with 15 groups in Excel (agile values, willingness to change, fixed structures, and hierarchies, agile leadership, technical vs. cultural agility, respect, transparency of decisions, transparency of processes, trust, perseverance, feedback culture, experiments, failure culture, comfort zone, and flight levels). This preliminary result was intensively discussed among the authors and then refined where appropriate.  

\paragraph{Step 3:} In the next step, we derived items for our quantitative questionnaire. We took our 15 codes from Step 2 and used them for this purpose. First, we formulated one problem statement for each code based on the corresponding data sets. Then we reformulated these problem statements into questions. These items were discussed and refined among the authors in several iterations. An example that illustrates this process is shown in Table~\ref{tab2:Example of derivation}.

\begin{table}
 \caption{Example of the derivation of a challenge (see Step 3)~\cite{Kuchel.2023}}
  \label{tab2:Example of derivation}
  \begin{tabular}{p{0.8cm}p{3.7cm}p{3.7cm}p{3.7cm}}
\hline
ID & Problem statement & Item (1. Iteration) & Item (Final) \\
\hline
C5 & The existing structures of an organization prevent an agile way of working. & How important is it to create structures that support an agile way of working? & How important is it to create organizational structures that support an agile way of working?\\
\hline
\end{tabular}
\end{table}

\paragraph{Step 4:} We then conducted the quantitative questionnaire study. Accordingly, we set up an online survey using Google Forms. We used appropriate guidelines for developing our questionnaire \cite{Graf.2002,Finstad.2010,Runeson.2009} and pretested it with five participants. In sum, the questionnaire comprises 21 items. Five of those items queried socio-demographic data which helps us to describe our sample, and 15 items queried the potential key challenges with agile culture (see C1-C15 in Appendix A.1) and one item posed additional comments. The participants of the survey rated the potential key challenges with agile culture (see C1-C15 in Appendix A.1) in terms of importance using a 7-point Likert scale (totally unimportant, unimportant, rather unimportant, neutral, rather important, important, and totally important). In addition, participants had the option to give no statement. 

\subsection{Data Collection and Analysis}
In the following, we present more details regarding the data collection and analysis of our quantitative questionnaire (see Step 4). The quantitative questionnaire was online between 11/26/2021 and 12/17/2021. It was spread via the professional network LinkedIn and personal contacts of the authors. In sum, we received 93 fully completed datasets. No participant terminated the questionnaire before the end. So, the dropout rate is 0\%. This surprised us, but can be explained by the good design of the questionnaire, oriented towards the target group. However, we removed one set of data from the sample because one participant showed unusual response behavior. The unusual response behavior was identified by analyzing the data set in detail. We found one respondent who selected the same value of the Likert scale for all questions except one.  

Furthermore, we used statistical data (mean, standard deviation, and confidence interval) for the analysis of the data. We defined challenges as key in those cases where 50\% of the participants weighted them as totally important. Based on this interpretation, we were able to identify a total of seven key challenges (see Table~\ref{tab3:Key Challenges with AC}).

\subsection{Description of the Sample}
The participants of the quantitative questionnaire (sample size N=92) had between 0 to 30 years (mean 6.13) of experience with an agile way of working. Of the 92 participants, 74 work in the private sector, 9 in the public sector, 10 work at universities or research institutes while 18 are self-employed (multiple answers were possible). A detailed description of the sample is given below in Table~\ref{tab2_1:Example of sample}.

\begin{table}[h]
\centering
\caption{Description of the sample of the quantitative questionnaire study (Step 4)}
\begin{tabularx}{\textwidth}{cXccc}
\hline
\textbf{ID} & \textbf{Item} & \textbf{Options} & \textbf{n} & \textbf{N} \\ 
\hline
\multirow{2}{*}{E1} & \multirow{2}{*}{Experience with agile ways of working} & Yes & 85 & \multirow{2}{*}{92} \\
 &  & No & 7 & \\
\hline
\multirow{4}{*}{E2} & \multirow{4}{*}{Years of experience with agile ways of working} & $<$ 3 years & 21 & \multirow{4}{*}{85} \\
 &  & 3-5 years & 18 &  \\
 &  & 6-9 years & 20 &  \\
 &  & $>$ 9 years & 26 &  \\
\hline
\multirow{4}{*}{E3} & \multirow{4}{*}{General opinion on agility} & Overrated & 2 & \multirow{4}{*}{92} \\
 &  & Neutral & 1 &  \\
 &  & Applied where useful & 58 &  \\
 &  & Urgently needed & 31 &  \\
\hline
\multirow{3}{*}{D1} & \multirow{3}{*}{Type of organization} & Private & 74 & \multirow{3}{*}{92} \\
 &  & Public Institutes & 19 &  \\
 &  & Self-employed & 18 &  \\
\hline
\multirow{5}{*}{D2} & \multirow{5}{*}{Industry} & Consulting & 20 & \multirow{5}{*}{92} \\
 &  & E-Commerce & 12 &  \\
 &  & R\&D & 7 &  \\
 &  & IT & 56 &  \\
 &  & Other & 22 &  \\
\hline
\end{tabularx}
\label{tab2_1:Example of sample}
\end{table}

Moreover, participants work in different industries such as IT, consulting, e-commerce, research and development, automobile, logistics, eHealth, insurance, energy, and social. We asked the participants for their general opinion on agility and they answered as follows: “overrated, a buzzword to me” (2,2\%), “should be applied where appropriate” (63\%), “urgently needed in today's world” (33,7\%), and “neutral” (1,1\%). Summarizing this information, we can observe that on the one hand, the sample has a good mix in terms of their experience with agile ways of working as well as working in different industries. On the other hand, the sample is not opposed to agile ways of working.

\section{Key Challenges with Agile Culture}
\label{Sec4:Results}
In this section, we answer our first research question \textit{RQ1: What are the key challenges with agile culture?} In doing so, we have identified seven key challenges related to agile culture, which are presented in Table~\ref{tab3:Key Challenges with AC}. To evaluate the responses, we used the Likert scale options to range from 1 (totally unimportant) to 7 (totally important). The challenges in Table~\ref{tab3:Key Challenges with AC} were rated as totally important by over 50\% of our survey participants. Additionally, all key challenges were found to be between important and totally important, as indicated by the small confidence interval. The statistical data for all queried challenges is provided in Appendix~\ref{Appendix}.

\begin{table}
 \caption{Key Challenges with Agile Culture~\cite{Kuchel.2023}}
  \label{tab3:Key Challenges with AC}
  \begin{tabular}{p{1cm}p{6cm}p{1cm}p{1cm}p{1.1cm}p{0.9cm}p{1cm}}
\hline
ID & Item (EN) & Mean & Stand. deviation & Confid. (p=0.05) & totally important & N \\
\hline
C1 & Humans in an organization do not treat each other with respect. & 6.70 & 0.81 & 0.17 & 80.4\% & 92\\
\hline
C2 & Management expects change from employees without embodying agile values themselves. & 6.43 & 1.17 & 0.24 & 69.6\% & 92\\
\hline
C3 & The organizational culture does not create a context for trusting interactions. & 6.60 & 0.63 & 0.13 & 67.4\% & 92\\
\hline
C4 & Humans in an organization are not allowed to make mistakes. & 6.49 & 0.75 & 0.15 & 60.9\% & 92\\
\hline
C5 & The existing structures of an organization prevent an agile way of working. & 6.40 & 0.96 & 0.20 & 59.8\% & 92\\
\hline
C6 & Agile cultural change does not occur at all levels (individual, team, management) of the organization. & 6.23 & 1.12 & 0.23 & 53.3\% & 90\\
\hline
C7 & Feedback is not valued in an organization. & 6.35 & 0.76 & 0.16 & 51.1\% & 92\\
\hline
\end{tabular}
\end{table}

Our analysis revealed that there are challenges related to embracing agile values in the organization, particularly with respect (C1) and trust (C3, C4, and C7). Furthermore, our data highlighted issues with existing organizational structures (C5) that hinder agile ways of working, as well as issues with cultural change at various levels within an organization (C6). 

\subsection{Discussion of the Key Challenges with Agile Culture}

\begin{figure}
\centering
\caption{Clustering of key challenges according to technical agility and cultural agility}
\label{fig:clusteringprocess}
\includegraphics[scale=0.35]{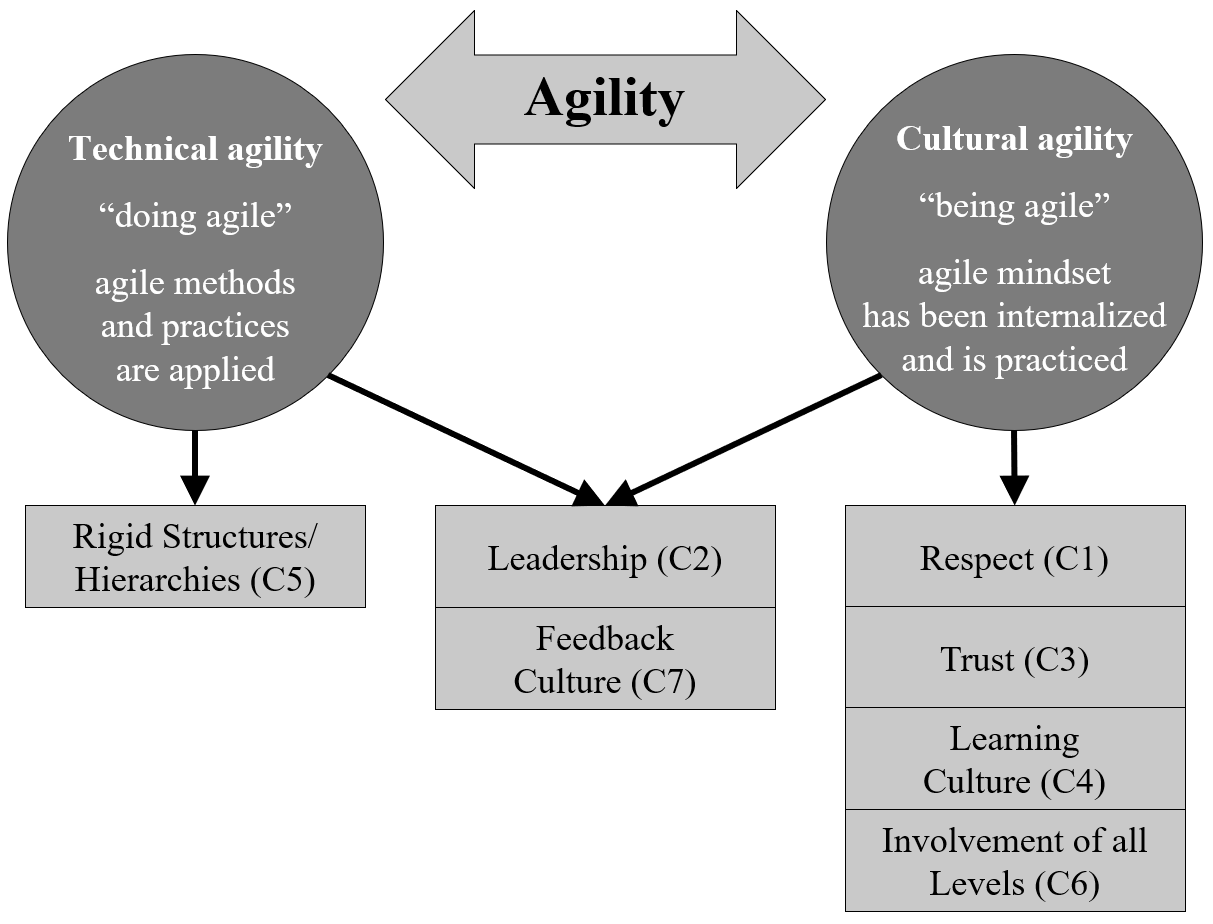}
\end{figure}

The identified key challenges with agile culture (see Table~\ref{tab3:Key Challenges with AC}) can be categorized into technical agility (doing agile) and cultural agility (being agile). Figure~\ref{fig:clusteringprocess} shows the result of the clustering process. The challenges leadership (C2) and feedback culture (C7) are borderline cases as parts of the problem relate to technical agility and other parts to cultural agility. In the case of leadership (C2, Management expects change from employees without embodying agile values themselves) parts such as existing control structures and micro management in the respective organization relates to technical agility, whereas embodying agile values or having a fixed mindset~\cite{Dweck.2017} is part of cultural agility. The case of feedback culture (C7, Feedback is not valued in an organization) relates to technical agility, e.g., if no tools or agile practices for providing feedback are in place, whereas valuing feedback is part of cultural agility.

In the following, we present a detailed explanation of each key challenge with agile culture.

Item \textbf{C1 (respect)} highlights that within an organization, there is a lack of respectful behavior towards each other, as indicated by 80\% of the respondents considering it very important. The workplace often faces the common issue of lacking respect in communication and interactions, which can be addressed to create a positive work environment. Fostering a respectful atmosphere promotes openness and transparency, empowering teams to self-manage their work. Agile leaders have a responsibility to exemplify respectful behavior, and when respect becomes a core value actively practiced within an organization, it can motivate employees and extend to the respectful treatment of customers. Mutual respect cultivates open communication and collaboration, fostering constructive and productive work processes.

Item \textbf{C2 (leadership)} represents another highly relevant aspect. The underlying leadership structure should emphasize agile values such as trust, respect, or open communication to support the continuous and step-wise adaption of agile practices in organizations. Today's complex work environment requires a different style of leadership to that of previous industrial revolutions. For example, agile leaders take responsibility and inspire performance. They also show empathy and vulnerability. Probably one of the biggest changes that modern organizations and leaders have to go through in the course of agile transformation is the delegation of responsibility to self-organized teams~\cite{Krieg.2022}. This change requires a shift in the mindset of leaders and an adaptation of existing control structures. 

Item \textbf{C3 (trust)} points to the significance of trust in an agile work culture. It emphasizes that trust among team members and between leaders and employees is crucial for cooperation and interdisciplinary work. The item reveals that 67\% (n=62) of respondents rated the absence of a framework for a trusting work environment as "very important." The lack of trust can impede the sharing of knowledge and skills, which are vital for applying agile methods and practices. Agile leaders should embody company values such as respect, courage, and openness to establish trust. A trusting work environment can eliminate time-consuming and costly control structures, allowing individuals to focus on their core tasks.

Item \textbf{C4 (learning culture)} is highly important within an organization in order to view mistakes as growth opportunities rather than solely blaming individuals. We propose that organizations should prioritize learning from mistakes instead of assigning blame. This can be achieved through open communication that encourages experimentation and public acknowledgment of mistakes. Agile leaders should lead by example by admitting and openly discussing their own mistakes. By fostering an environment where mistakes are not feared but embraced as chances for improvement, organizations can enhance innovation and motivation. It is crucial to emphasize that allowing individuals to make mistakes is insufficient; handling mistakes constructively and positively is equally vital.

Item \textbf{C5 (rigid structures/hierarchies)} points to the importance towards an agile way of working due to adapting the existing structures and hierarchies. A hierarchical culture has been shown to have a negative impact on the use of social agile practices (e.g. facilitating social interaction, collaboration, and direct communication) and a negative impact on the use of technical agile practices (coding/testing-oriented software engineering practices)~\cite{Gupta.2019}. For this reason, communication at eye level should be encouraged within an organization, regardless of the level (individual, team, management) at which it takes place.  

Item \textbf{C6 (involvement of all levels)} addresses the significance of a cultural shift in introducing agile working within an organization. The management level expects this change to initiate from the level below them, known as the flight levels (operational level, coordination, and strategic portfolio management)~\cite{Leopold2017}. However, a top-down approach cannot enforce a cultural change, and intrinsic motivation among employees at all levels is necessary for successful agile transformation. Agile leaders play a critical role by involving and engaging employees, helping them understand and identify with the change. Flight levels, which represent different perspectives on agile work within an organization, can be utilized to gradually introduce agile structures, without requiring every individual to work in an agile manner simultaneously. A cultural change on all levels, including leadership, is indispensable for long-term success in adapting to evolving organizational requirements.

Item \textbf{C7 (feedback culture)} is a pivotal aspect of agile work and influences organizational culture~\cite{Strode.2009}. Feedback enables continuous reflection and adaptation, while providing a safe space for individuals to express their thoughts and ideas. However, many organizations undervalue feedback. The quality of dialogue among colleagues and superiors influences relationship quality, which, in turn, reflects the company's culture. The retrospective, conducted at the end of an iteration, plays a key role in establishing a secure environment for a feedback culture. Feedback is recognized as the key to continuous development, lifelong learning, and constructive collaboration~\cite{Ram.2019}. Embracing an open feedback culture can help organizations better navigate ongoing environmental changes and transition towards agile work practices through continuous and constructive dialogue.

\section{Model of Agile Cultural Challenges}
\label{Sec5:Model}
\begin{figure}
\centering
\caption{The ACuCa conceptual model~\cite{Kuchel.2023}}
\label{fig:model}
\includegraphics[scale=0.28]{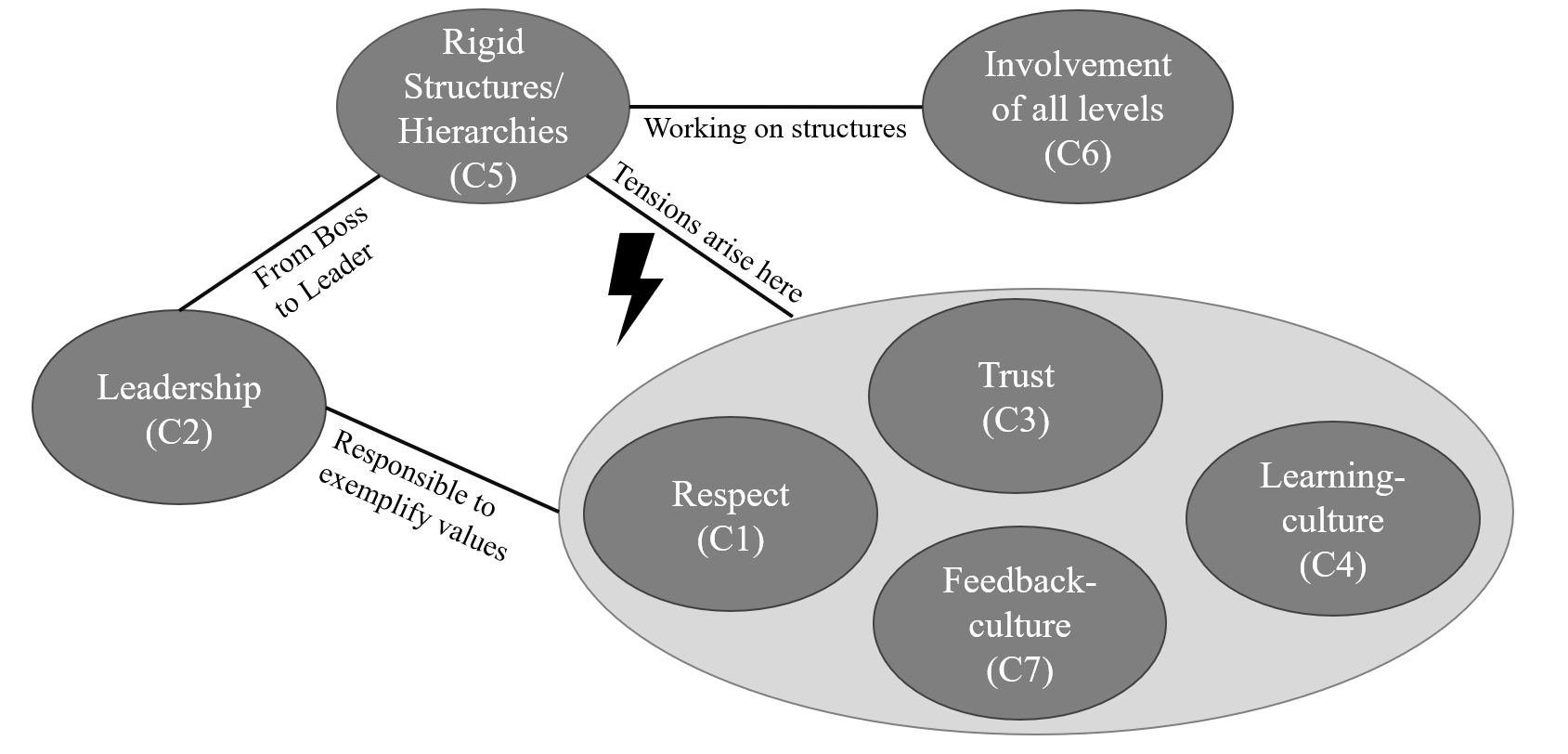}
\end{figure}

This section answers our second research question \textit{RQ2: How can we describe the relationships of the key challenges with agile culture in a systematic manner?} The clustering of the key challenges (see Fig. \ref{fig:clusteringprocess}) enables us to derive a conceptual model that shows the interrelationships between the key challenges with agile culture. A conceptual model allows us to clarify complex ideas and relationships by providing a visual representation. The result of the clustering process (see Fig.~\ref{fig:clusteringprocess})  provided the ground for creating our conceptual model named \textbf{A}gile \textbf{Cu}ltural \textbf{C}h\textbf{a}llenges (ACuCa) (see Fig. \ref{fig:model}). The ACuCa model helps to identify causal factors, dependencies and potential outcomes associated with the key challenges. In this Section, we first give a brief explanation about the creation process of the ACuCa conceptual model. Next, we introduce the model on detail and explain in particular the relationships among the clustered key challenges.

The model was created by the second author and improved based on a thorough review by the fourth author. The creation process consisted of two steps. In the first step, we clustered the key challenges. Our analysis led to the result that we have mainly different facets in the conceptual model. One facet covers the value-related key challenges, while the other facet consists of such challenges that implicitly or explicitly affect each other. Thus, we decided to create one cluster for the value-related key challenges C1, C3, C4, and C7. We also created three clusters, each consisting of one key challenge. 

The second step aimed to analyze and identify the realtionships among the key challenges using existing literature. We decided to use the guidelines of the well-known agile method Scrum~\cite{Schwaber.2020} as well as the Agile Manifesto~\cite{Beck.2001}. In addition, the analysis included the extracted insights from the identified related work. For example, our analysis led to an in-depth understanding of how specific key challenges (e.g. C5) influence the core of clustered value-related key challenges.

We found that fixed structures, established thought patterns, and strict hierarchies with a command-and-control approach are major barriers to agile ways of working. The existing structures of an organization hinder an agile way of working, and a change in both culture and structural conditions is required for introducing more agility. The interaction between the two dimensions of cultural and technical agility (see Fig.\ \ref{fig:clusteringprocess}) cannot be neglected, as structural conditions can have both performance-enhancing and inhibiting effects. Hierarchy, silos, and micromanagement pose a major hurdle to agile cultural development, and the biggest challenge lies in overcoming old paradigms and enabling a shift towards a structure that focuses on customer and business agility~\cite{Narayan2015}. In larger organizations, this can lead to the creation of silos that are more product-, value-, and customer-centric. It is important to adapt the dimensions of culture and structure in small steps that are coordinated with each other.

Organizational structures involve all levels (individual, team, organization), and organizations should iteratively adapt their structures to optimize interactions between levels while considering the organizational culture. Additionally, organizational structures contain the management level, and changing from a boss to a leader culture is a key factor for increased agility. This involves exemplifying values, behaviors, and a culture of open feedback and learning to establish a growth mindset~\cite{Dweck.2017} among employees. We define a boss as a manager who has a hierarchical understanding of the organization and a top-down approach to decision-making. We believe it is crucial for agile leaders to act as role models for agile teams within an organization.

To lead effectively in an agile environment, leaders must have experienced agility themselves and embody agile values. The role of a leader shifts towards becoming a coach and role model, rather than a commander. Self-organized teams require guidance through shared visions, values, and transparency, with a focus on communication and motivation. Leaders must set the frame for the team and plan adaptively while providing protection, trust, continuity, and support for employees' development through coaching and mentoring. By demonstrating and embodying agile values, leaders can facilitate the organization's adoption of agile practices step-by-step. The transformation towards an agile leadership style requires targeted training and development for future agile leaders.

\section{Discussion}
\label{Sec6:Discussion}
In the following section, we discuss the implications for practitioners and researchers and suggest ways in which the key challenges we have identified can help to promote an agile culture. In addition, we discuss the limitations of this study.

\subsection{Discussion of Implications for Practitioners and Researchers}
Comparing our research findings (see Section \ref{Sec4:Results}) to the existing literature (see Section \ref{Sec2:RelWork}), we discovered that cultural conflicts and resistance to change are major barriers to agile adoption~\cite{VersionOne.2023}. Our key challenges with agile culture (listed in Table~\ref{tab3:Key Challenges with AC}) emphasize the importance of respect (C1) and trust (C3, C4, and C7) as crucial factors in the success of an agile transformation. These challenges could be attributed to inadequate support and guidance during the transformation process~\cite{Kotter.1995}.

To address these challenges, our research provides practical solutions for both practitioners and researchers. Practitioners can use our key challenges as a tool to drive cultural change in their organizations. The challenges raise awareness of problems that come up when trying to integrate agility into an existing organizational culture, and help to prioritize which challenges are most important to tackle. For example, C6 highlights the importance of considering the pace of cultural change across different levels (individual, team, organization), and practitioners can use this information to develop appropriate solutions.

Changing organizational culture is a complex process that requires time and appropriate models or frameworks~\cite{Kotter.1995,Krieg.2022}. The Cynefin framework~\cite{Snowden.2007}, Haufe Quadrant~\cite{Haufe.2023} or the Deming circle~\cite{Deming.1986} are useful models for initiating and managing change. Researchers can support agile transformation and cultural change by developing appropriate tools, for example, to measure agility and thus support agile transformation~\cite{Looks.2021,Patel.2009,Siakas.2007}. Based on this tool support, researchers may develop new or adapted approaches, methods and even tools specifically focused on agile cultural change in an organization, using our identified key challenges as a basis for its development. Our key takeaways for practitioners and researchers are summarised below:
\newpage
\begin{framed}
\textbf{Key takeaways:} 
\begin{itemize}
    \item Live the agile values, especially respect and trust
    \item Leadership must embody agile values
    \item Trust is the basis for cooperation, knowledge transfer and interdisciplinary interaction
    \item Innovation requires experimentation, which sometimes fails 
    \item Focus on customers and users 
    \item Communication at eye level instead of power and hierarchy
    \item Transformation should take place at all levels (Flightlevels model)
    \item Agile working is about living and sharing in real time, so it needs a culture that values feedback
\end{itemize}
\end{framed}

\subsection{Threats to Validity \& Limitations}
Despite our rigorous adherence to establishing a research protocol by means of using appropriate guidelines and using a profound research approach, here are limitations and threats in our study that must be considered and discussed, according to Wohlin et al.\ ~\cite{Wohlin.2012}:

\textit{Construct validity:} Our mixed-method research approach included pretests and feedback from the target group to ensure the validity of our data collection. To prevent interpolations in our quantitative survey, we used a 7-point Likert scale and provided the option for participants to choose "no comment" for each item. As a lesson we learned, next time we would rephrase "no comment" to "no answer" even if from our perspective no comment fits better to our concrete question. Furthermore, for the comparison and check with the existing state of the art, we used a rapid review~\cite{Cartaxo2020}. We are aware of the fact that the replicability of the survey is limited due to the open target group of our professional network. Nevertheless, we did not only focus on agile practitioners in the network but spreaded it to all. However, we know that the practitioners in our community are more related to agile than other topics. Apart from a higher dropout rate or lower participation rate, we accepted this threat for our survey. 

\textit{Internal validity:} The design of the questionnaire is crucial for accurate data collection. Therefore, we conducted several pretests in different stages with individuals who met the criteria of the target group. All these pretests resulted in a 0\% dropout rate. In addition, we removed the data set of the one participant who showed conspicuous response behaviour.

\textit{External validity:} The surveys were conducted in German, as we aimed to reach individuals within our personal professional/business networks. While this approach allowed us to achieve a 0\% dropout rate in the quantitative survey, it may limit the generalizability of our results to other groups. However, since our demographics showed that many participants work in the international IT industry, the general transferability of our findings may not be significantly impacted. To assess the generalizability of our results further we have started further studies to conduct surveys in other languages.

\section{Conclusion and Future Work}
\label{Sec7:ConclusionAndFutureWork}
This paper is an extended version of our study \textit{Which Challenges Do Exist With Agile Culture in Practice?}~\cite{Kuchel.2023}. This extended version presents the findings of a study aimed at addressing the key challenges of implementing agile culture in practice. We further present ACuCa as a conceptual model explaining the relationships between the identified challenges. Seven key challenges, including respectful treatment between humans (C1), agile leadership (C2), trust in interactions (C3), learning culture (C4), rigid hierarchies (C5), involvement of all organizational levels (C6), and not valued feedback (C7), emerged from the interplay of organizational culture and agile practices. 

The ACuCa model reveals that a rigid hierarchical structure impedes the development of an agile culture aimed at optimizing aspects such as value-based work and learning culture. Moreover, involvement of all organizational levels, including management, teams, and individuals, is essential for optimizing both the structure and culture of the organization. To support agile software development teams' self-optimization, we encourage the management level to transform from a boss to a leader and exemplify relevant values for successful agile method usage. Based on our results, we provide practical implications and specific recommendations for cultural change in organizations, such as utilizing transformation frameworks like the Cynefin framework or the Deming circle. 

However, our findings reveal the need for further research on cultural influences on agile methods and practices. As agile methods become more tailored, selecting the right approach in specific situations becomes more complex. The various cultural aspects and influences described at different levels, such as the department or specific teams, may pose complex challenges in organizations. Therefore, we call on other researchers to study the relationship and interplay of cultural influences on the use and transition of agile methods in software development and beyond. We support the findings of other researchers who have suggested the usage of transformation frameworks and encourage more research to be conducted in this area, as there is still much to be explored.  

\appendix
\section{Appendices}
\subsection{Challenges with Agile Culture}
\label{Appendix}
The full data set including the 15 challenges with agile and organizational culture is shown in Table~\ref{tab4:Appendix-Key Challenges with AC}.
\begin{table*}
 \caption{Appendix 1: Key Challenges with Agile Culture~\cite{Kuchel.2023}}
  \label{tab4:Appendix-Key Challenges with AC}
  \begin{tabular}{cp{0.3\linewidth}cp{1cm}p{1cm}p{1cm}p{1cm}cc}
\hline
ID & Item (EN) & Mean & Stand. deviation & Confid. (p=0.05) & Confid. intervall min. & Confid. intervall max. & totally important & N \\
\hline
C1 & Humans in an organization do not treat each other with respect. & 6.70 & 0.81 & 0.17 & 6.53 & 6.86 & 80.4\% & 92\\
\hline
C2 & Management expects change from employees without embodying agile values themselves. & 6.43 & 1.17 & 0.24 & 6.20 & 6.67 & 69.6\% & 92\\
\hline
C3 & The organizational culture does not create a context for trusting interactions. & 6.60 & 0.63 & 0.13 & 6.47 & 6.73 & 67.4\% & 92\\
\hline
C4 & Humans in an organization are not allowed to make mistakes. & 6.49 & 0.75 & 0.15 & 6.34 & 6.64 & 60.9\% & 92\\
\hline
C5 & The existing structures of an organization prevent an agile way of working. & 6.40 & 0.96 & 0.20 & 6.21 & 6.60 & 59.8\% & 92\\
\hline
C6 & Agile cultural change does not occur at all levels (individual, team, management) of the organization. & 6.23 & 1.12 & 0.23 & 6.00 & 6.47 & 53.3\% & 90\\
\hline
C7 & Feedback is not valued in an organization. & 6.35 & 0.76 & 0.16 & 6.19 & 6.50 & 51.1\% & 92\\
\hline
C8 & Humans in an organization do not want to change the way they work. & 6.16 & 1.03 & 0.21 & 5.94 & 6.37 & 46.7\% & 90\\
\hline
C9 & In an organization, there is no transparency regarding decisions. & 6.35 & 0.76 & 0.16 & 6.13 & 6.44 & 45.7\% & 92\\
\hline
C10 & In an organization, there is no transparency regarding processes. & 6.25 & 0.76 & 0.16 & 6.09 & 6.41 & 45.1\% & 91\\
\hline
C11 & Humans in an organization do not understand that agile transformation needs perseverance. & 6.13 & 0.94 & 0.19 & 5.94 & 6.33 & 44.4\% & 90\\
\hline
C12 & Humans in an organization don't want to leave their comfort zone. & 5.97 & 1.05 & 0.22 & 5.75 & 6.18 & 37.8\% & 90\\
\hline
C13 & There is no safe environment for experiments in the organization. & 5.97 & 0.87 & 0.18 & 5.79 & 6.15 & 30.8\% & 91\\
\hline
C14 & Humans in an organization have not internalized the difference between technical and cultural agility. & 5.73 & 1.18 & 0.24 & 5.49 & 5.97 & 30.3\% & 89\\
\hline
C15 & The cultural values of humans within an organization do not align with agile values. & 5.96 & 1.12 & 0.23 & 5.72 & 6.19 & 30.0\% & 90\\
\hline
\end{tabular}
\end{table*}

\bibliographystyle{splncs03}
\bibliography{references}


%
%


\end{document}